\documentclass[aip,apl,amsmath,amssymb,reprint,superscriptaddress]{revtex4-1}

\usepackage[utf8]{inputenc}
\usepackage[T1]{fontenc}
\usepackage{graphicx}
\usepackage[dvipsnames]{xcolor}
\usepackage{dcolumn}
\usepackage{bm}
\usepackage[colorlinks=true, linkcolor=RoyalBlue, citecolor=RoyalBlue]{hyperref}
\usepackage[per-mode=symbol, separate-uncertainty]{siunitx}
\usepackage[capitalise]{cleveref}

\begin{document}

\title{Ultralong spin lifetimes in one-dimensional semiconductor nanowires} 
\author{Florian Dirnberger}
\affiliation{Institut für Experimentelle und Angewandte Physik, Universität Regensburg, D-93040 Regensburg, Germany}
\author{Michael Kammermeier}
\affiliation{Institut für Theoretische Physik, Universität Regensburg, D-93040 Regensburg, Germany}
\author{Jan K\"onig}
\affiliation{Institut für Experimentelle und Angewandte Physik, Universität Regensburg, D-93040 Regensburg, Germany}
\author{Moritz Forsch}
\affiliation{Institut für Experimentelle und Angewandte Physik, Universität Regensburg, D-93040 Regensburg, Germany}
\author{Paulo E. Faria Junior}
\affiliation{Institut für Theoretische Physik, Universität Regensburg, D-93040 Regensburg, Germany}
\author{Tiago Campos}
\affiliation{Institut für Theoretische Physik, Universität Regensburg, D-93040 Regensburg, Germany}
\author{Jaroslav Fabian}
\affiliation{Institut für Theoretische Physik, Universität Regensburg, D-93040 Regensburg, Germany}
\author{John Schliemann}
\affiliation{Institut für Theoretische Physik, Universität Regensburg, D-93040 Regensburg, Germany}
\author{Christian Schüller}
\affiliation{Institut für Experimentelle und Angewandte Physik, Universität Regensburg, D-93040 Regensburg, Germany}
\author{Tobias Korn}
\affiliation{Institut für Experimentelle und Angewandte Physik, Universität Regensburg, D-93040 Regensburg, Germany}
\author{Paul Wenk}
\affiliation{Institut für Theoretische Physik, Universität Regensburg, D-93040 Regensburg, Germany}
\author{Dominique Bougeard}
\email{*dominique.bougeard@ur.de}
\affiliation{Institut für Experimentelle und Angewandte Physik, Universität Regensburg, D-93040 Regensburg, Germany}

%\date{\today}

\begin{abstract}
We experimentally demonstrate ultralong spin lifetimes of electrons in the one-dimensional (1D) quantum limit of semiconductor nanowires. Optically probing single wires of different diameters reveals an increase in the spin relaxation time by orders of magnitude as the electrons become increasingly confined until only a single 1D subband is populated after thermalization. We find the observed spin lifetimes of more than \SI{200}{\ns} to result from the robustness of 1D electrons against major spin relaxation mechanisms, highlighting the promising potential of these wires for long-range transport of coherent spin information.\\
\end{abstract}

\maketitle

%Introduction
Nanowires (NWs) present three key assets: their unique shape, an exceptional surface-to-volume ratio and a high level of control during the epitaxial crystal growth. These features have established NWs in a cornerstone role for an impressively diverse area of nanoscale concepts, ranging from custom-tailored light-matter interaction \cite{Bleuse2011,Akimov2007,Oulton2009}, energy harvesting and sensing \cite{Garnett2010,Patolsky2005} to ballistic quantum transport \cite{Zhang2017}. By controlling the diameter at the nanoscale, NWs can for instance be tailored to a specific application by matching them with the length scale of a particular (quasi-) particle. Introducing radial spatial quantum confinement for electrons in semiconductor NWs, thus leaving only one direction of free motion, opens an experimental route to fascinating new phenomena such as Majorana-bound states \cite{Zhang2018}, the unique Coulomb interactions in Tomonaga-Luttinger liquids \cite{Deshpande2010}, unusual dispersion effects in spin-orbit coupled 1D systems \cite{Rashba2012} or long-range, coherent spin transport. Promising groundwork towards long-range spin transport has been demonstrated in wire-like, but yet diffusive systems \cite{Malshukov2000,Kiselev2000,Holleitner2006,Kettemann2007,Nitta2009,Wenk2010,Altmann2014,Kammermeier2016,Schwemmer2016,Kammermeier2017}. While these studies highlight a correlation between the wire width and the spin relaxation, going beyond diffusive systems by pushing experiments into the one-dimensional (1D) quantum limit should give access to a new realm of spin coherence.

In this Letter, we present a series of GaAs NWs with different diameters to investigate how spin relaxation evolves in the transition from a continuous three-dimensional (3D) dispersion to the electronic 1D quantum limit, where only a single 1D subband is occupied. Our optical approach allows us to investigate single, free-standing NWs. In our NW system, spatially confining electrons to 1D is expected to completely remove the usually very efficient mechanism of Dyakonov-Perel spin relaxation \cite{Dyakonov1972}. We indeed experimentally observe extraordinarily long spin relaxation times of more than \SI{200}{\ns} for the thinnest NWs: an increase by a factor $\sim\SI{500}{}$ for the transition from 3D to 1D. We demonstrate that the spin relaxation in our experiment is a result of the electron-hole (e-h) exchange interaction. Our analysis shows that the confinement of e-h pairs to increasingly smaller length scales very efficiently suppresses this exchange-driven spin relaxation in our NWs, causing the strong increase observed in our experiment, and illustrates the robustness of 1D electron spins against relaxation.

The GaAs/$\textrm{Al}_{0.36}\textrm{Ga}_{0.64}\textrm{As}$ core/shell NWs were synthesized in the wurtzite (WZ) phase with high phase purity \cite{Furthmeier2014} by molecular beam epitaxy using the Au-seeded vapour-liquid-solid growth technique on GaAs(111)B substrates. The NWs are nominally undoped and grow vertically on the substrate with the growth direction parallel to the WZ $\hat{c}\parallel\langle0001\rangle$-axis. By focusing the laser beam to a spot size of $\sim\SI{1}{\um}$, we were thus able to utilize confocal micro-photoluminescence ($\si{\micro}$-PL) spectroscopy to address single upright standing wires and investigate their emission properties \cite{Furthmeier2016}. 
All PL spectra were obtained at $\SI{4.2}{K}$ under continuous-wave (cw) or pulsed (\SIrange[range-units = single,range-phrase = --]{70}{500}{\ps} pulses at a repetition frequency of \SI{1}{\MHz}) excitation of a near-resonant (1.58eV) laser diode. The laser power is adjusted depending on the wire diameter (see supplemental material). The emitted PL was imaged onto the entrance slit of a grating spectrometer and detected by a charge-coupled device. Time-resolved photoluminescence signals were acquired by a Hamamatsu streak camera system with a minimum time resolution of \SI{\sim50}{\ps}.\\
	\begin{figure}
		\includegraphics{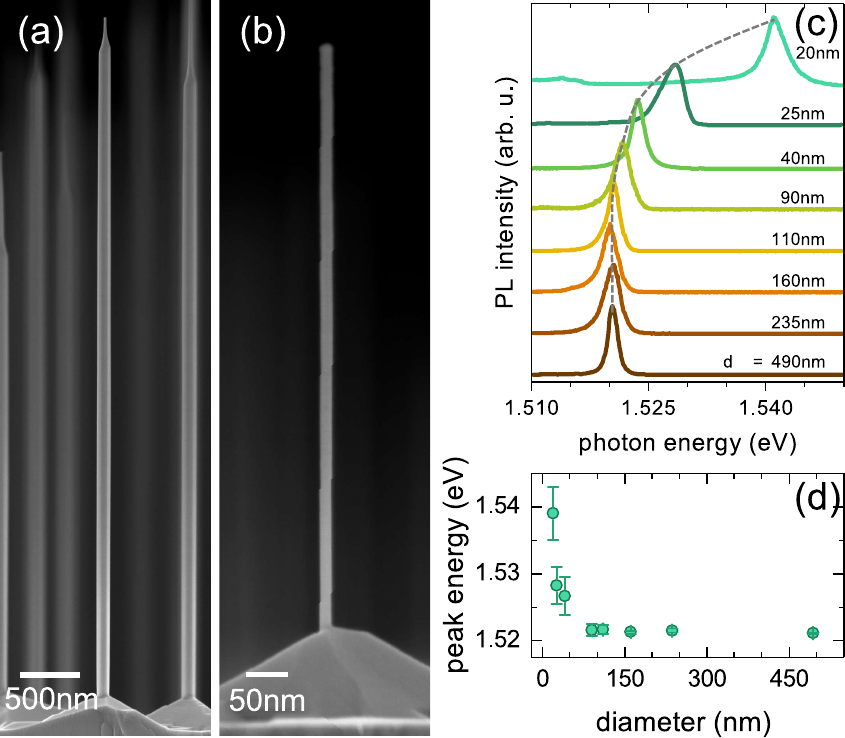}
		\caption{Side-view scanning electron micrographs of single, free-standing GaAs NWs with diameters (a)~$\textrm{d}=\SI{113}{nm}$ and (b)~$\textrm{d}=\SI{22}{nm}$. Note the different scale bars of both subfigures. (c)~Luminescence emission from NWs with different diameters. Each single NW spectrum represents one of the eight individual wafers produced for our series. Spectra were normalized for illustration purposes. The apparent broadening of the luminescence linewidth for smaller NW diameters is discussed in the supplementary material. (d)~The peak energy of the NW emission increases as a consequence of spatial quantum confinement. Each value is obtained by averaging the peak energy from several single NWs. The standard deviation (error bars) increases for smaller NWs, consistent with the statistical diameter distribution of a self-assembled ensemble. \label{fig:1}}
	\end{figure}
We fabricated eight individual wafers in total, each resulting in a NW ensemble with a narrow distribution (\SI{+-5}{\nm}) around the average diameters: \SIlist[list-units = single]{20;25;40;90;110;160;235;490}{\nm}. \Cref{fig:1} shows two exemplary scanning electron micrographs of such single free-standing GaAs NWs with the respective diameters of (a)~\SI{113}{\nm} and (b)~\SI{22}{\nm}. For all eight NW wafers the GaAs core was passivated by a \SI{10}{\nm} thick $\textrm{Al}_{0.36}\textrm{Ga}_{0.64}\textrm{As}$ shell to suppress non-radiative recombination at the bare GaAs surface. A \SI{5}{\nm} thick GaAs cap prevents oxidation of the shell. The length of the NWs decreases from $\sim\textrm{\SIrange[range-units = single]{6}{1}{\micro\meter}}$ as the diameter decreases. In the following, size indications refer to the diameter of the GaAs core as measured by scanning electron microscopy between two opposite corners of the hexagonal cross-section.

A general observation we make in our \si{\micro}-PL studies on single wires is that upon decreasing the NW diameter below \SI{50}{nm}, we observe a significant increase in the emission energy. To demonstrate this behaviour, we show a series of time-integrated \si{\micro}-PL spectra, obtained under cw excitation, for eight different single NWs, which are representative for the respective wafer, in \cref{fig:1}c. To compensate for the absorption losses in thin wires (see supplemental material), we increase the cw laser power from $\textrm{\SIrange[range-units = single]{0.03}{2.50}{\watt\per\square\centi\meter}}$. The averaged emission energy of several single NWs of nominally identical diameter is further summarized in \cref{fig:1}d for each of the eight different wafers. While no significant energy shift occurs for NW diameters in the range of \SIrange[range-units = single]{490}{90}{\nm}, a clear increase in the emission energy can be observed for $\textrm{d}<\SI{50}{\nm}$. For the larger NWs, we measure an average emission peak energy of \SI{1.521}{\eV}, which is consistent with earlier reports of the low-temperature PL emission energy in WZ GaAs NWs \cite{Furthmeier2014,Ahtapodov2012}. Upon decreasing the diameter below \SI{50}{nm}, the emission continuously shifts towards higher energies by a total amount of $\Delta E\approx\SI{20}{meV}$. We attribute this spectral shift to the increasing spatial quantum confinement in the NW core. It marks the transition from a continuous, 3D dispersion to a quantized 1D system in our thinner NWs \cite{Vainorius2016}. We estimate the splitting between the 1D subbands and find that, under the optical excitation used in our experiments, only the first subband is occupied by thermalized electrons in NWs with diameters smaller than $\SI{35}{\nm}$, while the same condition is fulfilled for the holes at diameters $\textrm{d}<\SI{90}{\nm}$ (see supplementary material). As a consequence, in our experiment, photoexcited electrons in wires with $\textrm{d}<\SI{35}{\nm}$ represent a 1D quantum system with only one populated subband. In the following, these wires will be denoted as 1D NWs. Covering the diameters from \SIrange[range-units = single]{490}{20}{\nm} thus allows us to study the evolution of the NW spin dynamics in the transition from a 3D dispersion to a true 1D system. 

We performed time- and polarization-resolved experiments, in which a single free-standing NW is excited with a circularly polarized ($\sigma_+$) laser pulse propagating parallel to the axis of the NW \cite{Furthmeier2016}. In the WZ crystal phase optical orientation thus creates an ensemble of e-h pairs which is homogeneously spin-polarized along the propagation direction of light \cite{Birman1959-1}. Upon recombination, these e-h pairs will emit partially polarized luminescence. Resolving the emission into left ($I_{-}$) and right ($I_{+}$) circularly polarized components directly links the experiment to the relaxation of the optically injected spin polarization. The spin relaxation time $\tau_{s}$ can then be extracted from fitting the degree of polarization $P_\text{C}=\left(I_+-I_-\right)/\left(I_++I_-\right)$ to a single exponential decay function, or by separately fitting the difference ($I_{+}-I_{-}$) and sum ($I_{+}+I_{-}$) signals (see supplementary material). With this optical orientation experiment, we have recently determined an effective g-factor in WZ GaAs NWs of $|g^*|=0.28$ \cite{Furthmeier2016}, which can be unambiguously attributed to conduction electrons, since the effective g-factor of heavy holes in this configuration is zero \cite{Broser1980,DeLuca2017}. 
Supported by our estimate suggesting rapid hole spin relaxation , this renders our experiment sensitive to the spin relaxation processes of the conduction electrons \cite{Hilton2002,Korn2010}. 
We estimate the e-h pair density $n$ under pulsed excitation. In the 3D NWs we find $n_{\mathrm{3D}}=\SI{8}{\times10^{16}{\cm}^{-3}}$, whereas $n_{\mathrm{1D}} \approx \SI{1}{\times10^6\cm^{-1}}$ in the 1D NWs. Such carrier densities characterize an e-h system slightly above the metal-insulator (or Mott-) transition--the notion of an exciton is therefore no longer strictly applicable in the pulsed excitation regime of our experiments \cite{Hildebrand1978,Zhai2012,Sarma2000}--and we thus use the term \textit{e-h pair}.\\
	\begin{figure}
		\includegraphics{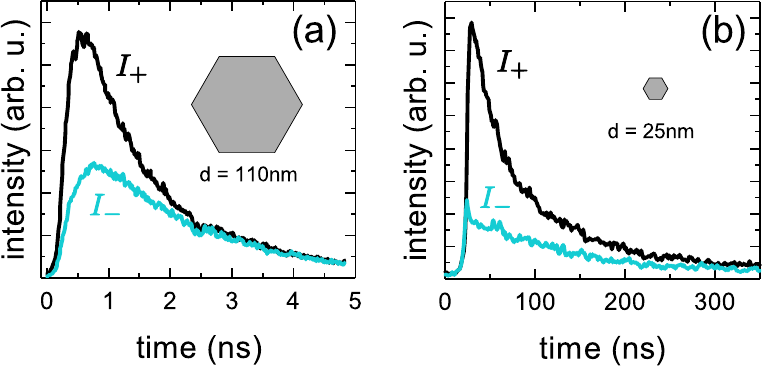}
		\caption{Time-resolved decay traces of the $I_{+}$ (black curve) and $I_{-}$ (blue curve) circularly polarized emission from two single NWs with the respective diameters (a) $\SI{110}{\nm}$ and (b) $\SI{25}{\nm}$. Note the different time scales in both measurements. The relative size of the two NWs is indicated by the inset. In (a), the difference between the two oppositely polarized traces decays within \SI{2.5}{\ns} after excitation. (b)~Even \SI{250}{\ns} after exitation, the two traces do not yet overlap. This already suggests a much longer timescale for spin relaxation in a very thin NW.  \label{fig:2}}
	\end{figure}
In \cref{fig:2}, we present two exemplary sets of decay traces, as obtained directly from the streak camera images. The curves show the temporal evolution of the polarized emission in a spectrally integrated narrow (\SI{5}{\meV}) window centered at the peak of the PL emission.  \Cref{fig:2}a shows the temporal decay of the circularly polarized emission of a 3D NW with a diameter of $\sim\SI{110}{\nm}$ on the scale of a few \SI{}{\ns}. We observe a large splitting between the $I_{+}$ and $I_{-}$ component, which decreases as a function of time until the two curves merge at $\sim\SI{2.5}{\ns}$. This time scale provides a rough measure of the spin relaxation. From our data analysis  we determine a spin relaxation time of $\SI{1.0}{\ns}$ and a photocarrier lifetime of $\SI{1.7}{\ns}$.

Opposed to this 3D case, we find the time scale on which spin relaxation occurs to be very different for the thin NWs. This is demonstrated for a 1D NW ($\textrm{d}=\SI{25}{\nm}$), as shown in \cref{fig:2}b. In this case, the splitting between the $I_{+}$ and $I_{-}$ curves decays over hundreds of \SI{}{\ns} with a photocarrier lifetime of \SI{87}{\ns} (see supplementary material). At $t>\SI{250}{\ns}$, the two curves do not yet appear to be in equilibrium, but their difference vanishes below the noise level. For the NW shown in \cref{fig:2}b, we determine a spin relaxation time of $\tau_{s}=\SI{98}{\ns}$. This strong increase of the spin relaxation time from $\SI{1}{\ns}$ in a wide 3D to $\SI{98}{\ns}$ in a narrow 1D NW already suggests the occurence of a strong suppression of the dominant spin relaxation mechanism.\\	
	\begin{figure}
		\includegraphics{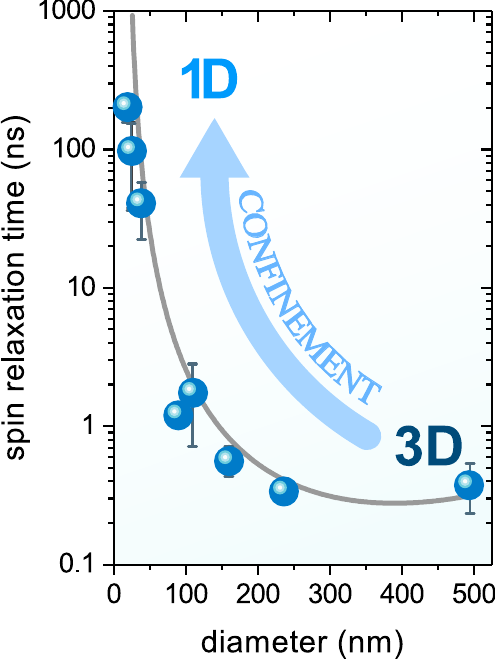}
		\caption{The spin relaxation time as a function of the NW diameter is displayed on a semilogarithmic scale. Blue symbols represent the average of measurements from several (3-8) single NWs and error bars indicate the statistical standard deviation. Spin relaxation times increase by more than two orders of magnitude as the NW diameter decreases from \SIrange[range-units = single]{490}{20}{\nm}. The diameter-dependence of the spin relaxation time as calculated from an exchange-based model  is plotted as a solid grey line. The model shows excellent agreement with the experimental values. The photo-excited e-h pair density was slightly above the Mott transition (see supplemental material). \label{fig:3}}
	\end{figure}
%
% Beschreibung des Graphen  
To map the evolution of spin relaxation in the transition from 3D to 1D, we have measured and determined the spin relaxation time for the full diameter range from \SIrange[range-units = single]{490}{20}{\nm}. The statistically averaged spin relaxation times of several (3-8) single NWs from each of the eight different wafers are summarized and displayed in \cref{fig:3}.

For the largest NWs of our study ($\textrm{d}=\SI{490}{\nm}$) we find relaxation times of $\tau_{s}=\SI{0.4}{\ns}$. 
By reducing the NW diameter in the experiment, we progressively confine the free carrier motion to a movement along the NW axis--a process that gradually induces a fundamental transition in the dimensionality of the electronic band structure. In the range from \SIrange[range-units = single]{235}{90}{\nm} we find only a weak increase of the spin relaxation time as the diameter decreases. Interestingly, entering the regime where the transition from 3D to 1D gets clearly visible in \cref{fig:1}c,d correlates with a \SI{35}{}-fold increase of the spin relaxation time between NW diameters \SIrange[range-units = single]{90}{40}{\nm}. When we then reduce the diameter beyond the 1D limit ($\textrm{d}<\SI{35}{nm}$), the increase becomes very steep, culminating in the observation of the spin relaxation time $\tau_{s}=\SI{202}{\ns}$ in the thinnest NWs investigated in our study: an increase by a factor of five hundred compared to the 3D NWs. Spin relaxation times of this order are extraordinarily long for mobile carriers in III-V semiconductors. 

%Discussion
Given the obvious and drastic weakening of the spin relaxation with increasing confinement through the NW diameter reduction, we will discuss a model for the spin relaxation in the following. 
%

%Dyakonov-Perel:
All previous studies of spin relaxation in diffusive, wire-like systems \cite{Malshukov2000,Kiselev2000,Holleitner2006,Kettemann2007,Nitta2009,Wenk2010,Altmann2014,Kammermeier2016,Schwemmer2016,Kammermeier2017} have focused on the most commonly observed spin relaxation mechanism in III-V semiconductors, the Dyakonov-Perel (DP) relaxation, based on a momentum-dependent spin splitting in the presence of spin-orbit coupling (SOC) \cite{Dyakonov1972,Zutic2004,Fabian2007}. Advantageous in this regard, for the NWs studied here, this spin splitting is intrinsically zero for electrons moving along the NW-axis due to the symmetry of the WZ crystal \cite{Rashba1959,*Rashba2015,Gmitra2016,Campos2018}. Confining the carrier motion more and more along this axis will thus ultimately eliminate the DP mechanism in the transition to the 1D NW regime, which we reach at $\textrm{d}=\SI{35}{\nm}$. Interestingly, we observe ultralong spin relaxation times for this diameter regime. Note, as a side remark, this absence of the DP mechanism also holds for any 1D NW in the zincblende phase growing along the cubic [111] direction \cite{Gmitra2016, Campos2018}, demonstrating its general relevance for III-V semiconductor NWs. 
In fact, not only the 1D NWs, but already our 3D NWs display a strongly reduced DP relaxation. Indeed, the relaxation time $\tau_{s}=\SI{0.4}{\ns}$, observed for the 3D NWs with the largest diameters ($\textrm{d}=\SI{490}{\nm}$), is already much longer than the few picoseconds predicted for these wire diameters in a DP model \cite{Kammermeier2018} assuming the fully diffusive carrier motion (see supplementary material). Notably, we thus have clear experimental evidence that the DP mechanism is strongly suppressed in the NWs and that all prerequisites of the diffusive regime are not fulfilled throughout the entire diameter range studied in our experiment. At the same time, \cref{fig:3} implies that the residual relaxation mechanism acting instead seems to be characterized by a strong NW diameter dependence and a likewise strong weakening when entering the 1D NW regime.

%Bir-Aronov-Pikus:

Our optical approach simultaneously creates an equal number of electrons and holes in the system, inducing exchange coupling between electron and hole spins. Taking this exchange coupling into account, Bir, Aronov and Pikus (BAP) developed a mechanism of spin relaxation, which critically depends on the exciton Bohr radius $a_{B}$: the spin relaxation time scales as $\tau_{s}\propto{a_{B}}^{-6}$\cite{Bir1975}. Note that, although our experiment rather probes the dynamics of free e-h pairs instead of excitons, the exciton Bohr radius is still a natural unit of length to describe the semiconductor system.
From the data discussed in \cref{fig:1}d, we have direct access to the exciton binding energies in our NWs, allowing us to deduce the corresponding exciton Bohr radius $a_{B}$ for each diameter value. The obtained diameter-dependence of $a_{B}(d)$ is accurately described by a polynomial function (see supplementary material). Introducing this $a_{B}(d)$ into the BAP spin relaxation $\tau_{s}(d)= \tau_{s,0}\; a_{B}(d)^{-6}$ leads to the solid line shown in \cref{fig:3}. Note that the values of $a_{B}(d)$ were determined completely independently of the spin lifetime experiments and that the solid line in \cref{fig:3} contains no fit parameters. It was only calibrated with $\tau_{s,0}$, the spin relaxation time of the largest NWs. \Cref{fig:3} reveals an excellent agreement between the calculated spin relaxation times (solid line) and the experimental data. Our analysis thus demonstrates that the electron spin relaxation time observed in our experiment throughout the transition from 3D to 1D is driven by the exchange-induced interaction between the photo-excited electrons and holes. Notably, it particularly captures the drastic increase of the spin relaxation times in the 1D NW regime, highlighting the efficient suppression of exchange-induced relaxation under increasing confinement.  

In III-V semiconductor structures, electron spin relaxation times of $\tau_{s}\ge\SI{100}{\ns}$ have so far only been observed for localized, 0D electrons \cite{Awschalom2001,Dzhioev2002,Hanson2003,Roemer2007}. In contrast to these 0D systems, in which the hyperfine (HF) interaction between carrier spins and fluctuating nuclear spins was found to limit the spin relaxation time, the delocalized wave function of 1D carriers effectively averages the fluctuating HF fields of many nuclei. The spin relaxation due to HF interaction is therefore much weaker in 1D as compared to 0D \cite{Khaetskii2002,Sousa2003}. Likewise, the importance of dopant ions in the mediation of dynamical nuclear polarization \cite{Salis2001,Huang2012,Lu2006} makes its occurrence highly unlikely. Since the 1D regime at the same time also strongly reduces the DP and BAP mechanisms, the final result is the observed exceptional robustness of 1D electron spins against relaxation. Furthermore, the electrons in our NWs are free to move along the 1D channel, making these wires ideal systems for the transport of coherent spin information, e.g., to interconnect spintronic devices on chip or allow coherent spin manipulation.

In conclusion, we have fabricated NWs which experimentally show a clear 1D character, i.e. only a single subband is occupied under photo-excitation. Entering this 1D NW regime is accompanied by the observation of spin relaxation times exceeding \SI{200}{\ns}, extraordinarily long for GaAs. We attribute the observation of such ultralong spin lifetimes to the complete suppression of the DP mechanism in III-V 1D NWs, a mechanism which is otherwise known to be highly efficient in most semiconductor structures. 
The small, but finite spin relaxation remaining in our 1D NWs is shown to be limited by the BAP mechanism. As the e-h pairs become increasingly confined in the 1D NWs, this mechanism is efficiently suppressed, resulting in more than 500 times longer spin relaxation times than in our larger NWs with a 3D dispersion. It is interesting to note that the spin relaxation limited by e-h exchange in our experiment is a consequence of the optical excitation, suggesting even longer spin relaxation times for pure 1D electron systems in both WZ and ZB III-V NWs.   

\section*{Supplementary Material}
See supplementary material for the calculation of the photoexcited carrier densities, further analysis of the 1D NW luminescence, the estimated 1D band dispersion, a detailed description of fitting the spin decay traces and an analysis of the Dyakonov-Perel and Bir-Aronov-Pikus mechanisms in nanowires.
%End Notes

\section*{Acknowledgements}
We thank M. Gmitra for helpful discussions and gratefully acknowledge financial support by the German Research Foundation (DFG) via SFB~689,~1277 and project 336985961. PEFJ and JF also acknowledge the financial support of the Alexander von Humboldt Foundation and Capes (grant No. 99999.000420/2016-06).
%
%References
%\section*{References}
\bibliographystyle{apsrev4-1}
%\bibliography{NWSpin_Manuscript}
%

%
%
\end{document}